\documentclass[twocolumn,a4,prl,preprintnumbers,amsmath,amssymb]{revtex4}

\usepackage{graphicx}% Include figure files
\usepackage{dcolumn}% Align table columns on decimal point
\usepackage{bm}% bold math

 \newcommand{\ket}[1]{\left|#1\right\rangle}

\begin{document}

%\preprint{AIP/123-QED}

\title{Sub-Kelvin optical thermometry of an electron reservoir coupled to a self-assembled InGaAs quantum dot}

\author{F. Seilmeier}
\author{M. Hauck}
\author{E. Schubert}
\author{G. J. Schinner}
\author{S. E. Beavan}
\email{sarah@beavan.com.au}
\author{A. H\"{o}gele}
\email{alexander.hoegele@lmu.de}

\affiliation{Fakult\"at f\"ur Physik and Center for
NanoScience (CeNS), Ludwig-Maximilians-Universit\"at M\"unchen,
Geschwister-Scholl-Platz 1, 80539 M\"unchen, Germany}

\date{\today}

\begin{abstract}
We show how resonant laser spectroscopy of the trion optical
transitions in a self-assembled quantum dot can be used to
determine the temperature of a nearby electron reservoir. At
finite magnetic field the spin-state occupation of the
Zeeman-split quantum dot electron ground states is governed by
thermalization with the electron reservoir via co-tunneling. With
resonant spectroscopy of the corresponding excited trion states we
map out the spin occupation as a function of magnetic field to
establish optical thermometry for the electron reservoir. We
demonstrate the implementation of the technique in the sub-Kelvin
temperature range where it is most sensitive, and where the
electron temperature is not necessarily given by the cryostat base
temperature.
\end{abstract}

\maketitle

Self-assembled semiconductor quantum dots (QDs) represent
promising building blocks for quantum information processing
\cite{Imamoglu1997}, and more recently have emerged as an
intriguing model system for optical studies of the quantum
impurity problem - the interaction of a localized electron with
the continuum of states in a fermionic reservoir
\cite{Tureci2011}. In the regime of strong tunnel coupling of a
resident QD electron to the nearby Fermi sea and sufficiently low
temperatures, signatures of many-body phenomena are observable in
emission \cite{Kleemans2010} or absorption with power-law tails
characteristic of the Fermi-edge singularity \cite{Haupt2013b} and
the Kondo effect \cite{Latta2011} in resonant spectra of neutral
and singly charged QDs. In addition to resonant laser spectroscopy
of charge-tunable QDs \cite{Hogele2004} and the control of their
exchange coupling to the Fermi reservoir enabled by the gate
voltage in QD field-effect devices \cite{Smith2005}, related
experiments crucially require cryogenic temperatures deep in the
sub-Kelvin regime \cite{Latta2011,Haupt2013b}.

While the temperature of the electron reservoir is a key parameter
in exploiting many-body phenomena, it is not necessarily the same
as that of the cryogenic bath, and is difficult to access
directly. In this Letter, we present a spectroscopic method to
determine the electron bath temperature. Our technique exploits
the sensitivity of spin-selective optical absorption in singly
charged QDs \cite{Hogele2005,Atature2006,Gerardot2008b} to
temperature. A measurement of the effective QD electron spin
temperature can be directly related to the spin bath temperature
of the Fermi reservoir \cite{Dreiser2008,Kroner2008a,Latta2011a}.
Although the QD-bath temperature relationship is complicated by
optical spin pumping (OSP), in the limit of strong exchange
coupling between the QD spin and the Fermi bath via co-tunneling,
the OSP is negligible, and the QD spin state occupation is
entirely governed by the thermal distribution of the electrons in
the Fermi sea. In either case (with or without OSP), the QD
electron spin polarization measured as a function of an external
magnetic field provides a direct measure of the electron bath
temperature.

In our experiment we used self-assembled InGaAs quantum dots grown
by molecular beam epitaxy \cite{Leonard1993} with intermediate
annealing \cite{Garcia1997}. The QDs were embedded inside a field
effect device \cite{Drexler1994} where a 25 nm thick GaAs
tunneling barrier separates the QDs from a heavily $n^+$ doped
GaAs layer that forms the Fermi reservoir. The QD-layer was capped
subsequently by $10$~nm GaAs, an AlGaAs/GaAs superlattice of
$252$~nm thickness, and $14$~nm of GaAs. A semitransparent NiCr
layer of $5$~nm was evaporated on the surface to form the top
electrode. A gate voltage applied to the top electrode tunes the
QD energy levels relative to the Fermi level pinned in the back
reservoir to control the QD charge occupation \cite{Warburton2000}
and the exciton emission energy through the quantum confined Stark
effect \cite{Warburton2002}. Moreover, the gate voltage also
varies the coupling between the QD electron spin and the
Fermi reservoir (given by the co-tunneling rate) by orders of
magnitude \cite{Smith2005}.

\begin{figure}[t]
\begin{center}
\includegraphics[width=8.5cm]{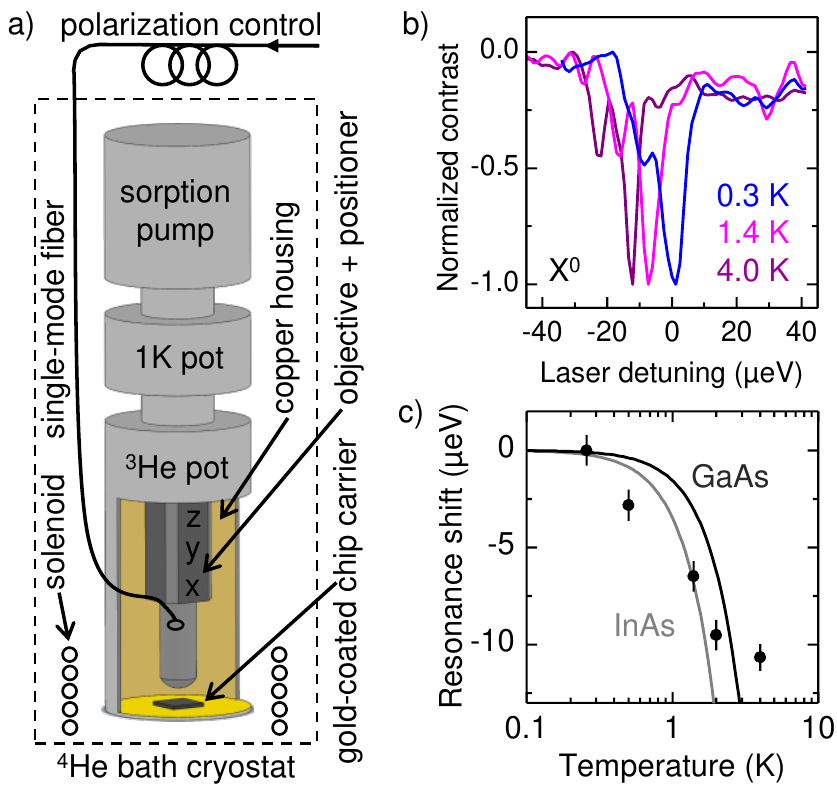}
\caption{(a) Experimental setup. The cryogenic system consists of
a $^3$He insert in a $^4$He bath cryostat and provides a minimum
base temperature of $250$~mK. The solenoid is used to apply
magnetic fields up to $10$~T along the vertical axis of the
cryostat. The quantum dot sample is mounted on a gold-coated chip
carrier in thermal contact with the $^3$He pot. Optical access to
individual quantum dots is enabled with a fiber-based
micro-objective mounted on an xyz nano-positionioner. (b)
Differential transmission spectra of the neutral exciton ($X^0$)
transition at cryostat base temperatures $T_{\text{base}}=0.3$~K,
$1.4$~K and $4.0$~K show a resonance red-shift with increasing
temperature. (c) The $X^0$ resonance shift as a function of
$T_{\text{base}}$. The grey and black lines represent the
temperature shift expected in bulk InAs and GaAs
respectively.}\label{Fig1label}
\end{center}
\end{figure}

The sample was mounted inside a $^3$He refrigerator with a nominal
minimum base temperature of $T_{\textrm{base}} = 250 \textrm{~mK}$
(Fig.~\ref{Fig1label}a). The temperature was adjusted from
$250$~mK to $4.0$~K by heating or pumping via the sorption pump on
the $^3$He pot. Optical access to the sample was provided by a
fiber-based confocal microscope system with a spot size of
$\approx 1~\mu$m \cite{Hogele2008}, addressing sufficiently few
dots for single QD spectroscopy. We used the differential
transmission method to address the neutral exciton ($X^0$) and
trion ($X^{-}$) optical transitions in a single QD with resonant
laser spectroscopy \cite{Hogele2004}.

The evolution of the $X^0$ resonance with temperature is shown in
Fig.~\ref{Fig1label}b and c. Both fine-structure resonances of the
neutral exciton exhibit a red-shift with increasing temperature
(data points in Fig.~\ref{Fig1label}b) consistent with a decrease
of the band-gap energy in semiconductors described by the Varshni
relation \cite{Varshni1967} (solid lines for bulk InAs and GaAs in
Fig.~\ref{Fig1label}b). The discrepancy between the measured
resonance shift and the expected bulk values is not surprising
given the uncertainty in both the material composition and the
strain distribution inherent to self-assembled QDs. It also
highlights the fact that a measurement of the resonance shift
alone does not qualify as a reliable method for quantitative
thermometry.

\begin{figure}[t]
\begin{center}
\includegraphics[scale=1.0]{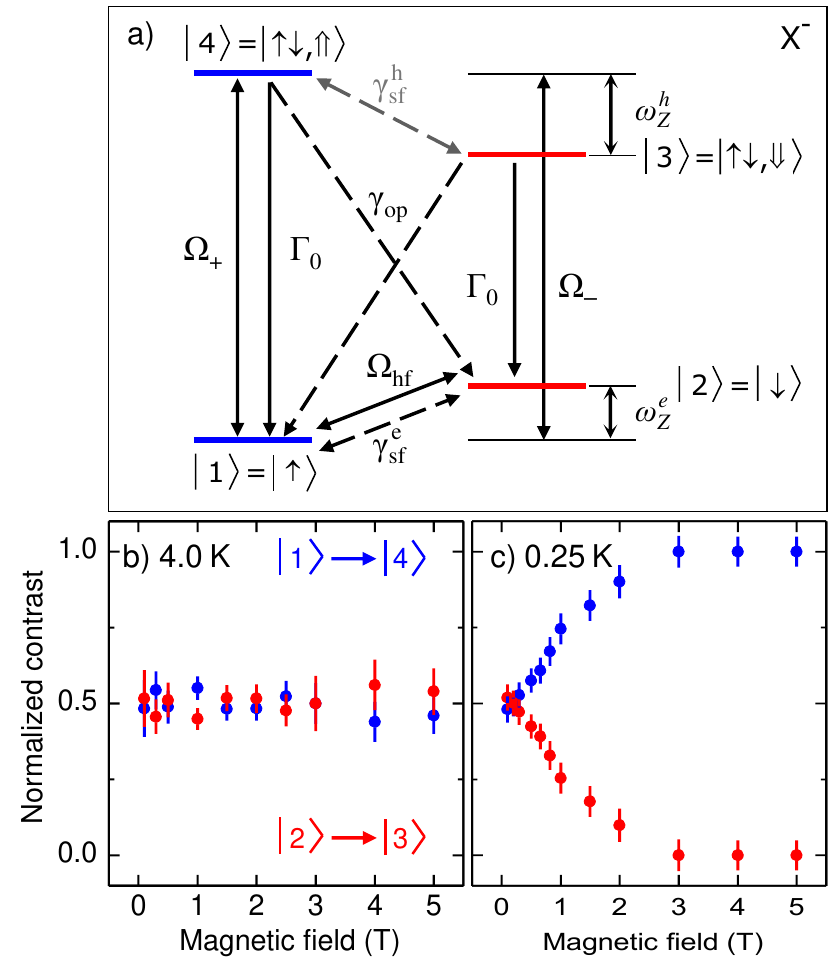}
\caption{(a) The four-level system associated with the $X^-$
resonance in a charged QD, with a magnetic field applied in
Faraday configuration. There are two dipole-allowed optical
transitions associated with the upper $|1\rangle
\leftrightarrow|4\rangle$ (blue transition) and lower $|2\rangle
\leftrightarrow |3\rangle$ (red transition) Zeeman branches. The
description of rates and Zeeman splittings is given in the text.
(b) and (c) Normalized contrast for both the red and blue 
transitions measured as a function of magnetic field with
$T_{\text{base}}=4.0$~K and $250$~mK, respectively. At large
magnetic fields and a sufficiently low temperature the spin
population accumulates in the spin-up ground state.
}\label{Fig2label}
\end{center}
\end{figure}

Instead, we exploit the temperature dependence of the
spin-resolved trion optical transitions in finite magnetic field
\cite{Hogele2005,Atature2006} to determine the electron bath
temperature. The level diagram of the $X^-$ in the presence of an
optical drive at finite magnetic field applied in Faraday geometry
is shown in Fig.~\ref{Fig2label}a. The lower electron states with
spin $\pm \frac{1}{2}$ (denoted as $\ket{\uparrow}=\ket{1}$ and
$\ket{\downarrow}=\ket{2}$, and split by the electron Zeeman
energy $\hbar \omega_{\mathrm{Z}}^{\mathrm{e}}=g_{\mathrm{e}} \mu_{\mathrm{B}} B$) couple to the trion states
with two spin-singlet electrons and one heavy-hole with spin $\pm
\frac{3}{2}$ (denoted as $\ket{\uparrow \downarrow \,
\Uparrow}=\ket{4}$ and $\ket{\uparrow \downarrow \,
\Downarrow}=\ket{3}$, and separated by the hole
Zeeman energy $\hbar \omega_{\mathrm{Z}}^{\mathrm{h}}=g_{\mathrm{h}} \mu_{\mathrm{B}} B$).  The dipole-allowed
transitions between the blue ($\ket{1}-\ket{4}$) and red
($\ket{2}-\ket{3}$) Zeeman branches can be selectively addressed
by $\sigma^+$ and $\sigma^-$ circularly polarized laser fields
with respective Rabi frequencies $\Omega_+$ and $\Omega_-$.

For strongly-confining QDs, the radiative decay rate $\Gamma_0$ is
insensitive to magnetic field for realistic experimental field
strengths and thus can be treated as equivalent for both
dipole-allowed transitions. It also provides an upper bound on the
dipole-forbidden diagonal transition rates
$\gamma_{\mathrm{op}}=\eta \Gamma_0$ that become weakly allowed by
heavy-hole light-hole admixing with $\eta \ll 1$ and contribute to
OSP \cite{Dreiser2008}. Although there is a temperature-dependence
in the incoherent hole spin-flip rate $\gamma_{\mathrm{sf}}^{\mathrm{h}}$
\cite{Gerardot2008b,Brunner2009,Houel2012}, this effect is
rendered negligible by the much faster optical decay channel, i.e.
$\gamma_{\mathrm{op}} \gg \gamma_{\mathrm{sf}}^{\mathrm{h}}$. Another temperature
insensitive parameter is the coherent coupling of the electron
spin-states mediated by the hyperfine interaction with a `frozen'
nuclear spin environment. This leads to an effective coupling
$\hbar\Omega_{\mathrm{hf}}\sim 1~\mu$eV \cite{Bracker2005,
Braun2005,Dreiser2008}, while the analogous coherent coupling of
the excited states is negligible due to much weaker hole hyperfine
interaction \cite{Fallahi2010,Chekhovich2010}. Finally, since both
the hyperfine-mediated and the spin-orbit induced spin-flip
processes are negligible in our experiment as compared to the
spin-exchange rate with the Fermi reservoir via co-tunneling (at a
rate $\gamma_{\mathrm{ct}}$), our inspection of the optically
driven four-level system arrives at the conclusion that the only
sensitivity to temperature stems from
$\gamma_{\mathrm{sf}}^{\mathrm{e}}=\gamma_{\mathrm{ct}}$. Importantly, the asymmetry
between the ground state occupations in the absence of OSP has an
exponential dependance on the temperature
$\gamma_{12}=\gamma_{21}\exp(-\omega_{\mathrm{Z}}^{\mathrm{e}}/k_{\mathrm{B}}T)$. This fact is
exploited in the following to determine the electron bath
temperature $T_\text{e}$ with trion laser spectroscopy.

\begin{figure}[t,b]
\begin{center}
\includegraphics[width=8.5cm]{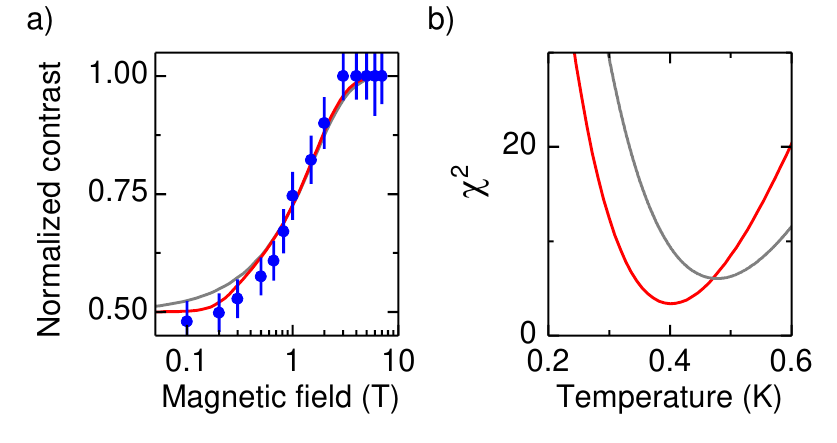}
\caption{(a) Normalized contrast measured at
$T_\text{base}=250$~mK, along with the temperature-fit results
using the full four-level model described in the text (red line,
$T_{\mathrm{e}}=400$~K), and the two-level thermal model (grey line,
$T_{\mathrm{e}}=480$~K). (b) The mean-squared `distance' between the data and
the model predictions for the normalized contrast indicates a
better fit by the four-level model (red line) as compared to the
two-level thermal distribution (grey line). }\label{Fig3label}
\end{center}
\end{figure}

Experimentally, it is convenient to use linear polarization to
address both trion transitions with one laser field scanned in
frequency, and we chose $\Omega_+ = \Omega_- \simeq \Gamma_0$ to
drive the transition close to saturation where the signal-to-noise
ratio of the differential transmission contrast $\alpha$ is
optimal \cite{Gerardot2007b}. Fig.~\ref{Fig2label}b and c
summarize the results obtained for the spin-resolved trion
branches at different magnetic fields and temperatures. For finite
magnetic fields, the two optical transitions were well resolved,
and the peak amplitudes were used to calculate the normalized
transmission contrast as $\alpha_{\text{blue,red}} / \left(
\alpha_\text{blue} + \alpha_\text{red}\right)$ for the blue and
red transition accordingly.

\begin{figure}[t]
\begin{center}
\includegraphics[width=8.5cm]{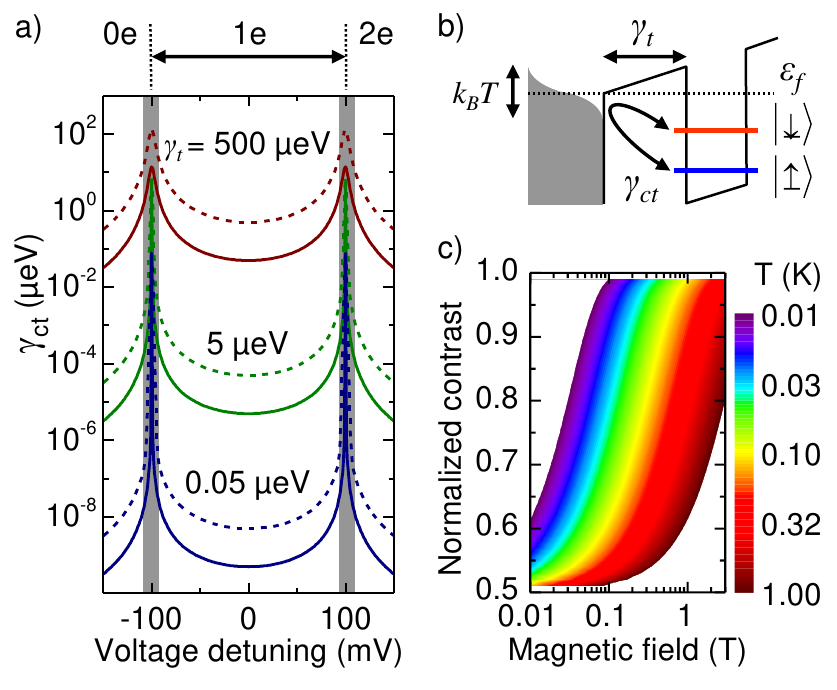}
\caption{(a) Optimal working regions for optical thermometry of
the electron reservoir at the edges of the charge stability
plateau are shaded grey. The co-tunneling rate
$\gamma_{\mathrm{ct}}$ is plotted as a function of gate voltage
detuning from the plateau center for 400~mK and 40~mK as dashed
and solid lines according to Ref.~\cite{Smith2005}. (b) At the
edges, the co-tunneling rate between the quantum dot electron
states and the Fermi edge $\varepsilon_{\mathrm{F}}$ thermally broadened by
$k_{\mathrm{B}} T$ is maximum for a sample-specific tunnel coupling
$\gamma_{\mathrm{t}}$ ($500~\mu$eV, $5~\mu$eV and $0.05~\mu$eV roughly
correspond to tunneling barriers of $15$~nm, $25$~nm, and
$35$~nm). Tuning to the maximum $\gamma_{\mathrm{ct}}$ simplifies
the system dynamics to an effective thermal two-level system, and
a single measurement of the normalized contrast at a particular
applied magnetic field should suffice to determine the temperature
of the electron reservoir using the color-map shown in (c).
}\label{Fig4label}
\end{center}
\end{figure}

The normalized contrasts in Fig.~\ref{Fig2label}b and c correspond
to nominal base temperatures of $4.0$~K and $250$~mK,
respectively. While there is no significant evolution of the
normalized contrast with magnetic field at $4.0$~K, the relative
strength of the blue transition grows at the expense of the red
transition for the lowest temperature of our $^3$He system. In
this case, the normalized contrasts saturate for magnetic fields
above $3$~T, implying a negligible population of the state
$\ket{\downarrow}$ and a spin accumulation in the $\ket{\uparrow
}$ state. This asymptotic limit is expected for a thermal spin
distribution in a singly charged QD governed by fast co-tunneling
processes \cite{Hogele2005}. At moderate magnetic fields, however,
the spin-state population is modified by optical spin-pumping
\cite{Atature2006} whenever $\gamma_{\mathrm{ct}} \simeq
\gamma_{\mathrm{op}}$. In our sample with a nominal separation of
$25$~nm between the electron reservoir and the QD-layer, we
estimate the tunnel coupling $\gamma_t$ (see
Fig.~\ref{Fig4label}b) in the range between $10~\mu$eV and
$50~\mu$eV for strongly confining QDs with emission around
$1.3$~eV. In the center of the trion stability plateau, the
working point in our experiments, this implies a competition
between effective thermal and optical spin-pumping rates at deep
sub-Kelvin temperatures, necessitating a full four-level system
analysis.

The four-level system is modelled using a Lindblad master
equation, similarly to Ref.~\cite{Gerardot2008b}.  The Hamiltonian
contains the coherent dynamics due to both optical fields
$\Omega_+$ and $\Omega_-$ with $\Omega_{\pm} = 2.5 \times
\Gamma_0$, and the hyperfine term
$\hbar\Omega_{\mathrm{hf}}=1.3~\mu$eV. The incident laser field
also drives the weakly allowed off-diagonal transitions with Rabi
frequencies of $\eta\Omega_{\pm}$, where we take $\eta \approx 4
\times 10^{-4}$ \cite{Dreiser2008}. The incoherent transition
rates; $\hbar\Gamma_0=1~\mu \textrm{eV}$,
$\gamma_{\text{op}}=\eta\Gamma_0$,
$\gamma_{21}=\gamma_{\text{ct}}$, and $\gamma_{12}$ as defined
above are included in the usual Lindblad superoperator formalism.
The value of the co-tunneling rate at the minimum base temperature
of 250~mK is estimated as $\hbar\gamma_{\mathrm{ct}}\approx
5\times 10^{-4}~\mu \textrm{eV}$ \cite{Smith2005}. An additional
term is included to account for the broadening of the optical
resonance that is caused by environmental charge and spin
fluctuations \cite{Hogele2004,Kuhlmann2013a}; in our experiments
the observed resonance width was $\hbar\Gamma \approx 6~\mu
\textrm{eV}$ (Fig.~\ref{Fig1label}b). We include the effect as
pure dephasing of the excited states with a rate of
$\Gamma_{\mathrm{d}}/2=2.5~\mu \textrm{eV}$ that contributes to the
experimental linewidth as $\Gamma=\Gamma_0+\Gamma_{\mathrm{d}}$. The electron
and hole $g$-factors are taken as $g_{\mathrm{e}} = 0.69$ and $g_{\mathrm{h}} = 0.81$
\cite{Kroner2008a}. The steady-state solutions for the density
matrix are found numerically \cite{Johansson2013}, for both cases
when the red and blue transitions are driven resonantly. The
normalized absorption contrast is calculated using the relevant
coherence terms of the density matrix.

The model is used to fit the data recorded at
$T_{\textrm{base}}=250$~mK with the temperature as the only free
parameter. The optimized fit gives a value of $T_\text{e} = 400\pm
50$~mK.  The solution of the four-level model is shown in
Fig.~\ref{Fig3label}a, along with the result expected from a
thermal population distribution between the two ground states. As
expected for this low value of $\gamma_{\mathrm{ct}}$ in our
sample in the sub-Kelvin regime, there is evidence of optically
induced spin pumping at low magnetic field values ($<500$~mT),
where the normalized contrasts remain closer to 0.5 (i.e. equal
population in both states $\ket{1}$ and $\ket{2}$) than would be
expected in a purely thermodynamic equilibrium.  The four-level
model qualitatively captures this spin pumping trend, and
therefore provides a better fit as compared to the two-level model
(see Fig.~\ref{Fig3label}b).  The temperature determined by the
four-level model fit changes by less then 10\% with variations in
the values of $\eta$ and $\Omega_{\mathrm{hf}}$ by a factor of 2,
or an order of magnitude change in $\gamma_{\mathrm{ct}}$. In
contrast, the model is more sensitive to the values of $\Gamma_0$,
$\Omega_{\pm}$ and $g_{\mathrm{e,h}}$.

For a general application of this thermometry method in alternate
QD systems, it would be desirable to eliminate this OSP signature,
and recover a simple two-level thermal system.  We suggest two straightforward alterations that would allow for this simplification.
Firstly, the effectiveness
of optically induced spin pumping could be reduced by pumping with
circularly rather than linearly polarized light such that only one
of the optical transitions is driven efficiently at small magnetic
fields.  Secondly, the relaxation rate $\gamma_{\mathrm{ct}}$ can be increased relative to
the optical spin-pumping channel, given by rates $\Omega_{\pm}$
and $\gamma_{\mathrm{op}}$.  The value of $\gamma_{\mathrm{ct}}$
can be controllably tuned across a few orders of magnitude by
varying the gate-voltage, and can be further altered for different
samples by tailoring the tunnel barrier energy itself.
Fig.~\ref{Fig4label}a shows how the co-tunneling rate varies with
gate voltage for a number of different tunnel-barrier energies,
and also for different temperatures \cite{Smith2005}.
It will in most cases be possible
to increase $\gamma_{\mathrm{ct}}$ sufficiently by tuning the gate
voltage (shaded regions in Fig.~\ref{Fig4label}a), and move into an elegantly simple regime where the
normalized contrast directly reflects a thermal distribution in a
two-level system.

Another mechanism through which the system dynamics will
significantly deviate from that of a two-level thermal
distribution could arise due to the dynamic interaction between
the electron spin and the $10{^5}$ nuclear spins in the QD
\cite{Urbaszek2013}.  The effect known as dragging (anti-dragging)
occurs when the electron-spin causes the nuclear-spins to align in
such a way as to Zeeman-shift the transition into (out of)
resonance with the incident light \cite{Hogele2012,Yang2012}. This
effect is particularly pronounced at high magnetic fields, long
integration times, and when the step-size of the laser frequency
sweep is small.  In the current experiment, dragging effects were
minimized by choosing a large laser step-size.  An alternative
method to eliminate nuclear spin magnetization would be to
actively depolarize the nuclear spin ensemble
\cite{Chekhovich2012c}.

If the experimental conditions are chosen such that the two-level
approximation is valid, then the electron reservoir temperature
$T_{\text{e}}$ can be read out with a single normalized-contrast
measurement using the colormap shown in Fig.~\ref{Fig4label}c. 
To quantify the precision of this method in determining the
temperature, we use the best-case signal to noise ratios
achievable with either differential transmission measurements
($6\times 10^3$ using a GaAs solid
immersion lens \cite{Vamivakas2007}) or resonance fluorescence
($10^5$, \cite{Kuhlmann2013a}).  This results in a temperature measurement with uncertainty 
as low as 0.004 \% using resonance fluorescence, or 0.06 \% with differential
transmission spectroscopy.  Therefore this method could potentially measure mK temperatures
with sub-$\mu$K precision.

In conclusion, we have developed and demonstrated a novel
technique to determine the temperature of an electron spin bath.
This was achieved by optically driving a QD tunnel-coupled to the
electron spin bath, and monitoring its spin polarization as a
function of magnetic field.  At mK temperatures, the actual
temperature of the electron reservoir can deviate significantly
from the nominal base temperature of the cryostat, as was the case
here where the temperature was determined as $400 \pm 50$~mK at a
cryostat base temperature of 250~mK.  By maximizing the
co-tunneling rate $\gamma_{\mathrm{ct}}$ via the gate-voltage, the
four-level system description simplifies to a two-level thermal
system, allowing for a straightforward and simple method of
electron bath thermometry.  The sensitivity of this method is
optimal at low temperatures ($T<1$~K).  The parameter range of low
temperatures and large co-tunneling rates coincides with the
regime of interest for exploring the many-body interactions
between a QD and an electron bath, where the reservoir temperature
is an important parameter. Thus, the optical thermometry technique
will be a particularly useful tool in future investigations of
many-body phenomena in self-assembled QD systems.

We acknowledge A. Badolato and P. M. Petroff for growing the
sample heterostructure used in this work, and M. Kroner for useful
discussions. This research was funded by the Deutsche
Forschungsgemeinschaft (SFB 631 and the German Excellence
Initiative via the Nanosystems Initiative Munich, NIM) with
support from the Center for NanoScience (CeNS) and LMUexcellent.

%\bibliography{library_v3}

\end{document}